\documentclass[conference]{IEEEtran}
\IEEEoverridecommandlockouts
\usepackage{cite}
\usepackage{amsmath,amssymb,amsfonts}
\usepackage{algorithmic}
\usepackage{graphicx}
\usepackage{textcomp}
\usepackage{xcolor}
\usepackage{subfig}
\usepackage{xcolor}
\usepackage{comment}
\usepackage{url}
\usepackage{hyperref}

\def\BibTeX{{\rm B\kern-.05em{\sc i\kern-.025em b}\kern-.08em
    T\kern-.1667em\lower.7ex\hbox{E}\kern-.125emX}}
\begin{document}

\title{Visualizing the Shadows: Unveiling Data Poisoning Behaviors in Federated Learning 
}


 \author{\IEEEauthorblockN{Xueqing Zhang\IEEEauthorrefmark{1}, Junkai Zhang\IEEEauthorrefmark{2}, Ka-Ho Chow\IEEEauthorrefmark{3}, Juntao Chen\IEEEauthorrefmark{1}, Ying Mao\IEEEauthorrefmark{1},  Mohamed Rahouti\IEEEauthorrefmark{1}, \\
 Xiang Li\IEEEauthorrefmark{1}, Yuchen Liu\IEEEauthorrefmark{4}, Wenqi Wei\IEEEauthorrefmark{1}\IEEEauthorrefmark{5}\thanks{\IEEEauthorrefmark{5}The corresponding author thanks the Fordham Faculty Research Grant.}}
\IEEEauthorblockA{\IEEEauthorrefmark{1} Department of Computer and Information Sciences, Fordham University \\ }
 \IEEEauthorblockA{\IEEEauthorrefmark{2}  Department of Applied Analytics, Columbia University \\ }
 \IEEEauthorblockA{\IEEEauthorrefmark{3}  Department of Computer Science, the University of Hong Kong \\ }
 \IEEEauthorblockA{\IEEEauthorrefmark{4} Department of Computer Science, North Carolina State University}
}

\maketitle

\begin{abstract}
This demo paper examines the susceptibility of Federated Learning (FL) systems to targeted data poisoning attacks, presenting a novel system for visualizing and mitigating such threats. 
We simulate targeted data poisoning attacks via label flipping and analyze the impact on model performance, employing a five-component system that includes Simulation and Data Generation, Data Collection and Upload, User-friendly Interface, Analysis and Insight, and Advisory System. Observations from three demo modules: label manipulation, attack timing, and malicious attack availability, and two analysis
components: utility and analytical behavior of local model updates highlight the risks to system integrity and
offer insight into the resilience of FL systems. The demo is available at~\url{https://github.com/CathyXueqingZhang/DataPoisoningVis}. 
\end{abstract}


\begin{figure*}[t]
    \centering
    \includegraphics[width=0.89\linewidth]{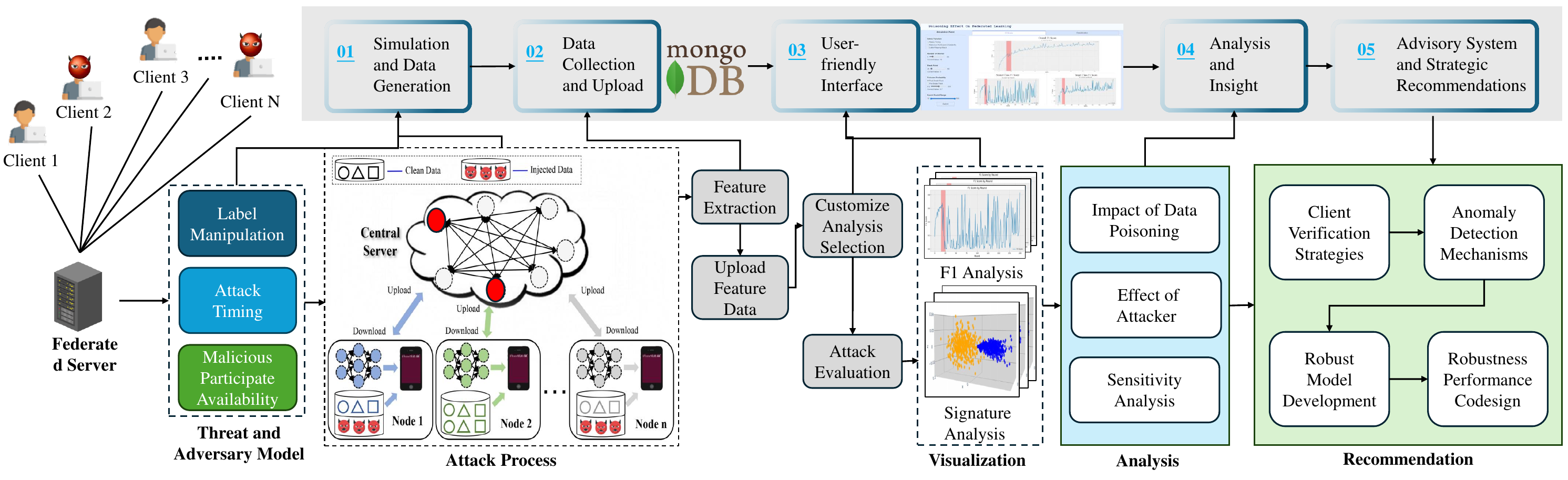}
    \caption{\small Overview.}
    \label{fig:Overview}
    \vspace{-0.4cm}
\end{figure*}

\section{Introduction}
Federated Learning (FL) is a promising machine learning framework that allows distributed clients to collaboratively train a model while keeping their data localized. Each client trains a model on its own data and sends the model updates to the server, which then aggregates these updates to improve a global model for the next round. This method effectively addresses the major privacy concerns of traditional centralized models. With the growing adoption of AI across various industries, FL's capability to harness collective without compromising individual privacy stands out as a significant advantage~\cite{mcmahan2017communication}. However, this distributed nature introduces unique vulnerabilities through data poisoning attacks where malicious agents can tamper the training process to compromise the model's accuracy performance or specific misbehaviors~\cite{biggio2012poisoning}.

Malicious participants in FL can exploit the lack of control over the clients by tampering with their local datasets or model updates. Errors will propagate through the federated network~\cite{biggio2012poisoning,tolpegin2020data} for model manipulation. Unlike the untargeted poisoning attacks that compromise the overall model performance for Denial-of-Service, 
the targeted data poisoning attack is particularly stealthy~\cite{chow2023stdlens,tran2018spectral}.
The latter modifies only specific inputs related to chosen victims, leaving the remainder of the model functioning as usual. 
Therefore, the targeted data poisoning attack poses 
significant risks to the integrity and reliability of the federated model~\cite{yuan2024robust,hayase2021spectre}. Highlighting these concerns emphasizes the need for robust defense mechanisms to ensure FL is trustworthy and effective.

In this demo paper, we present an empirical investigation into the vulnerability of FL systems to targeted data poisoning attacks, highlighting through visualizing the multifaceted landscape of the threats. Specifically, Our study delves into the extent to which FL systems are susceptible to these malicious activities and introduces a novel advisory system designed to detect and mitigate such threats. 
Our analysis sheds light on the countermeasure design, providing insight into robustness strategies. There are five components in the proposed demo system for visualizing targeted data poisoning attacks and presenting the adversarial behavior of the targeted data poisoning attack. 
The first component is a simulation of the targeted data poisoning attack via label flipping~\cite{tolpegin2020data}. We explore the effects of different attack timings, and malicious participant percentage and availability. The second component acts as the server and collects clients' local model updates for analysis. The experimental data are recorded in JSON format and stored on MongoDB. The third component consists of an interactive visualization tool, providing a user-friendly interface for analytical inspection of the local model update shared with the server from the clients. The fourth component aims to offer analytical insights from the behavior of the local model update, and the analysis of the impact of poisoning on F1 scores, the effectiveness of attacker classification, the sensitivity of different FL models to poisoning, and the extent of model drift due to poisoned inputs. Finally, the fifth component will help 
translating such analytical insights into actionable recommendations for robust FL system design.

With extensive experiments, our demo system generates valuable insights for defending against targeted data poisoning attacks. For example, 
our findings suggest that while the overall accuracy of the global model can rebound from attacks that occur early in the training process, attacks in later stages—particularly those with significant participation from malicious users—tend to cause more severe damage. In general,
we advocate for rigorous client verification strategies to preclude the participation of malicious agents. Based on the separable model updates from the benign and malicious clients, 
we recommend studying further when the incorporation of anomaly detection techniques can help identify and neutralize attacks promptly. 
Additionally, we emphasize the importance of designing robust FL models that maintain their performance in the face of such adversarial challenges, a concept we term "robustness performance co-design."




\section{Demo System Design}

The design overview of our demo system is illustrated in
\textbf{Figure~\ref{fig:Overview}}. By simulating various attack scenarios, including 
different types of label manipulation, attack timing, and malicious participant percentage and availability ~\cite{tolpegin2020data,wei2023demystifying}, the goal is to uncover the malicious behavior of the targeted data poisoning attacks on the shared local models and its impact on attack effectiveness. Our framework integrates an interactive dashboard facilitating the generation, feature extraction, and visualization of the FL model behavior in the presence of malicious clients. We focus on studying the effectiveness of the poisoning attack and the separable phenomenon of local model updates between the benign and the malicious clients.

\noindent\textbf{Simulations and Data Generation.}
The demo begins by simulating an FL environment with a central server and multiple clients (i.e., 50) over 200 global training rounds\footnote{\url{https://github.com/git-disl/DataPoisoning_FL.git}}. We consider 10\% participation rate, i.e., 5 participants are selected per round. We implement the label-flipping attack to poison the training dataset at local clients for its effectiveness and readiness to corrupt the FL learning process. Note that other types of targeted data poisoning attacks, e.g., backdoor, demonstrate a similar phenomenon in terms of victim data compromise and separable local model update between the benign and malicious clients~\cite{tran2018spectral,tolpegin2020data,steinhardt2017certified}. The users are given the choice to specify the victim class and the attack target class. Accordingly, malicious clients will modify the label of their training data from the victim class to the target class, causing misclassification. For each of the victim-target pair, we design three attack modules: Label Manipulation, Attack Timing, and Malicious Participant Availability. In all three modules, users have the flexibility to set the number of malicious workers, thereby adjusting the proportion of compromised participants in the FL network.

In \textit{Label Manipulation}, we evaluate the robustness against adversarial label manipulations using two benchmark datasets: CIFAR-10 and Fashion MNIST. For CIFAR-10, our experiments include switching labels from 'dog' to 'cat' (5 to 3), 'airplane' to 'bird' (0 to 2), 'automobile' to 'truck' (1 to 9), and 'deer' to 'horse' (4 to 7). These label changes are designed to test the FL system's ability to handle misclassifications between visually or contextually similar categories. In parallel, with Fashion MNIST, we manipulate labels from 't-shirt' to 'shirt' (0 to 6), 'trouser' to 'dress' (1 to 3), 'coat' to 'pullover' (4 to 2), and 'sneaker' to 'ankle boot' (7 to 9), which examines the framework's response to mislabeling among different types of apparel. 
In \textit{Attack Timing}, we investigate how attacks that occur only during certain phases of the training process can affect system performance. The default scenario uses the CIFAR-10 dataset with labels flipped from "automobile" (class 1) to "truck" (class 9). Users can also specify the training rounds during which the malicious workers are active, opting for scenarios where the attack commences after a certain point or ceases before another. 
Additionally, we examine the impact of \textit{Malicious Participant Availability}. The default scenario also uses the CIFAR-10 dataset with labels flipped from "automobile" (class 1) to "truck" (class 9). 
Given that the client selection process in FL usually relies on the availability of the client~\cite{mcmahan2017communication}, the attacker can have the motivation to always make themselves online. Thus, even though the percentage of the attackers among 50 clients can be as low as 10\%, the number of attackers selected at each round for global aggregation can be much higher. For instance, setting a parameter to 0.7 means there is a 70\% probability that any selected participant will be one of the malicious actors in any given round.

\begin{figure*}[t]
    \centering
    \subfloat[User-friendly Interface-F1 Analysis\label{fig:f1}]{%
        \includegraphics[width=0.45\linewidth,height=4.70cm]{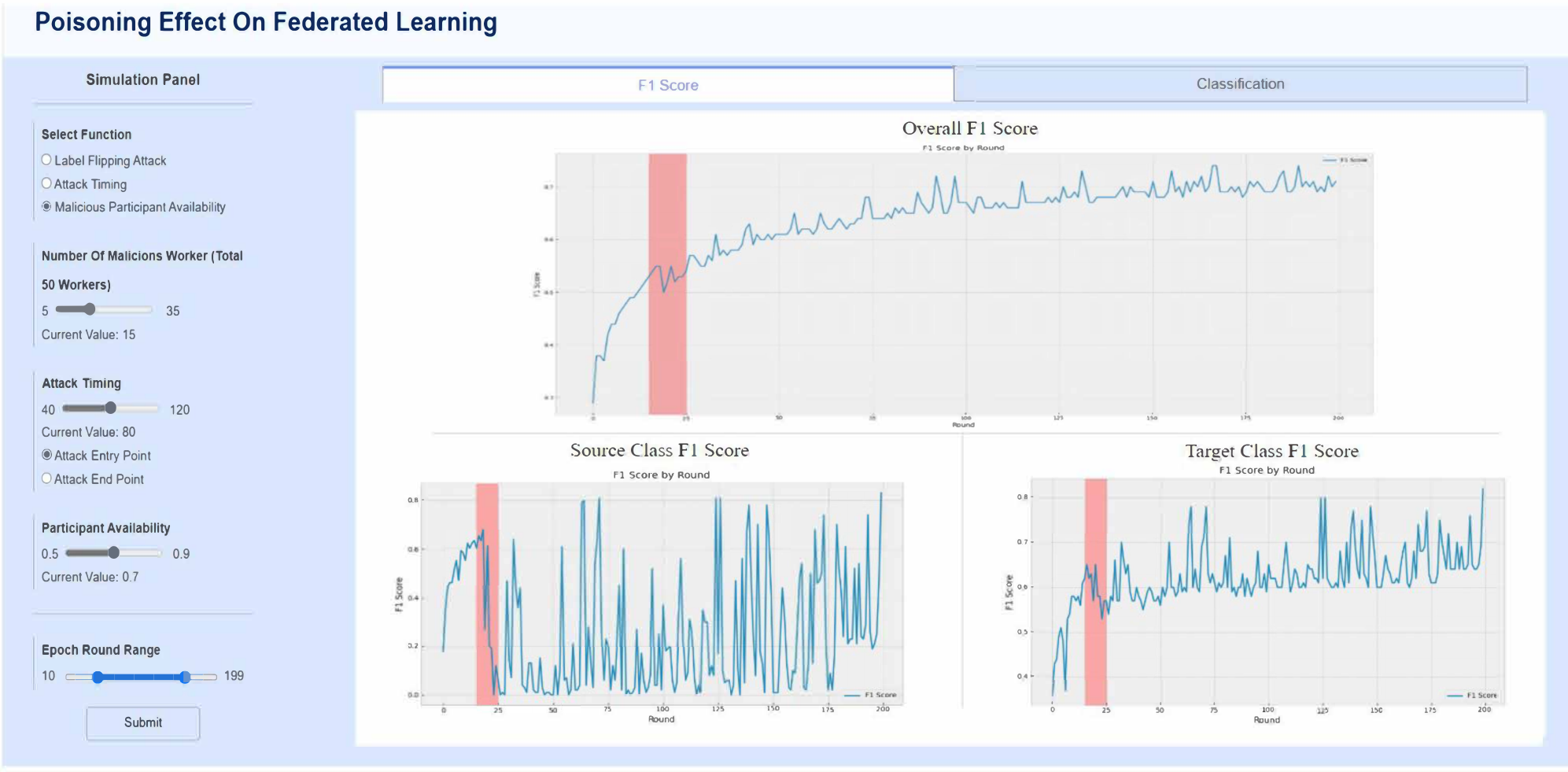}  \hspace{1.0cm}}
    \subfloat[User-friendly Interface-Signature Analysis\label{fig:pca}]
    {%
        \includegraphics[width=0.45\linewidth, height=4.70cm]{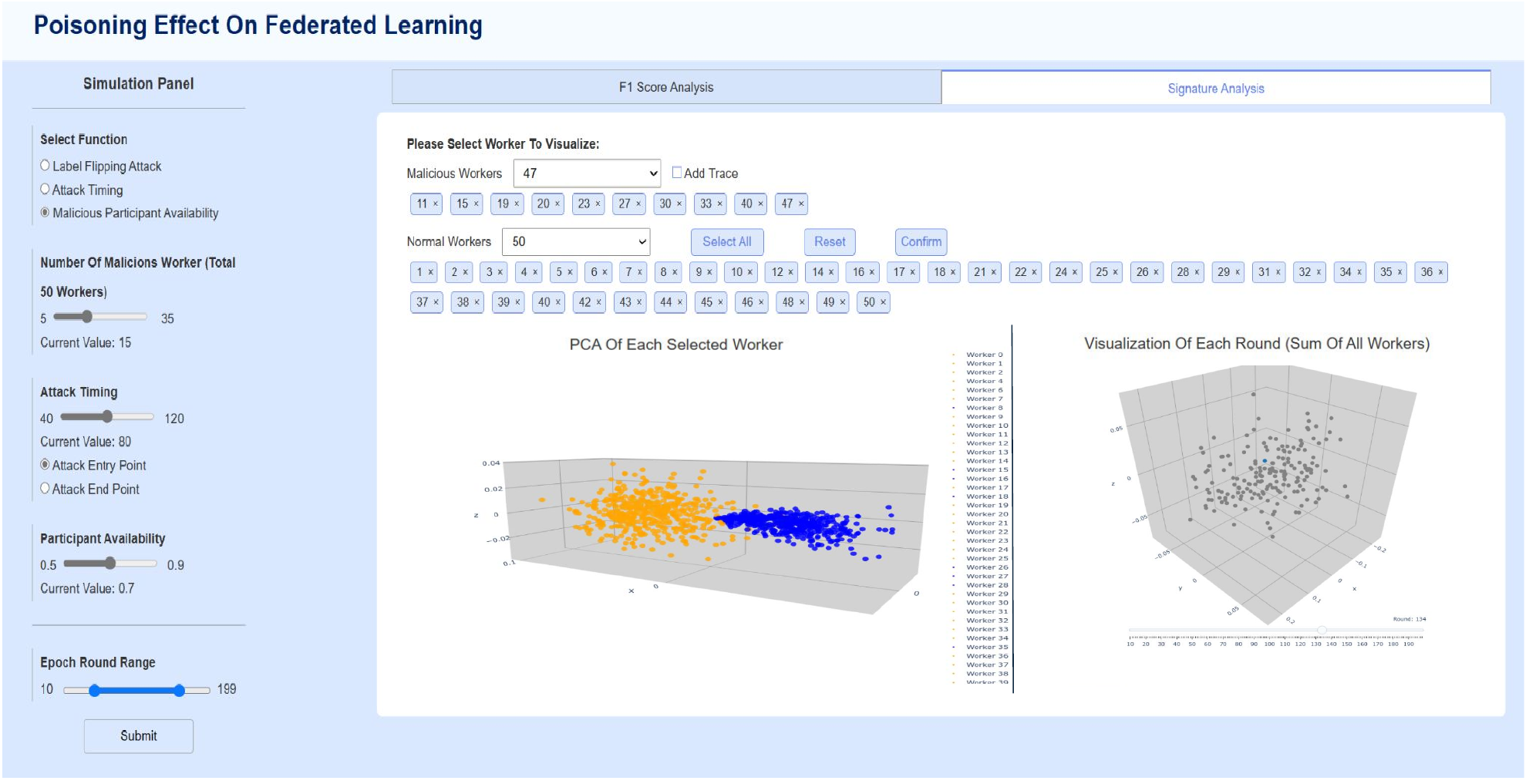}}
    \caption{\small User-friendly Interface}
    \label{fig:User-friendly Interface}
    \vspace{-0.4cm}
\end{figure*}

\noindent\textbf{Data Collection and Management.}
With local model update data collected at the global server throughout the FL learning process, a comprehensive dataset including client participation logs and round-specific model parameters is gathered and saved in JSON format. 
This dataset is stored at MongoDB as the backend database.







\noindent\textbf{User-friendly Interface.} An interactive dashboard is designed to allow users to engage with the visualization data dynamically. Users can scrutinize the effects of data poisoning, the performance of individual clients, and the progression of the model's learning over time. Our demo system features two aspects of the FL system in the presence of the label flipping attack: f1 score regarding the global model performance, and the three-dimension projection of the client's high-dimension local model updates under principal component analysis (PCA). In the first visualization aspect, 
we plot the F1 score of the global model under the user-specified attack setting will be used to study the global model performance throughout the course of FL learning, as shown in \textbf{Figure~\ref{fig:f1}}. In this module, the user will be able to zoom in on the F1 score for both the source victim class and the attack target class at the specified epoch round range. 

In the second component of our signature analysis, we generate three-dimensional PCA visualizations for the client's local model updates during each training round, as shown in \textbf{Figure~\ref{fig:pca}}. Users are informed about the identities of malicious and benign clients, allowing them to select specific rounds and clients for detailed spatial and temporal examination of the local model updates. Our approach to reducing the high-dimensional model updates to three dimensions is inspired by the observed decoupling effect between poisoned and benign updates as documented in existing literature~\cite{tran2018spectral,hayase2021spectre,chow2023stdlens}. These visualizations facilitate an understanding of the learning dynamics and identify the rounds in which model performance begins to deteriorate due to poisoning. Additionally, we offer a separate 3D PCA visualization that aggregates local model updates from all five workers at each epoch round. As the round progresses, each point is iteratively highlighted in blue, illustrating how gradient updates evolve over time. 

\section{Observation and Insights}
\begin{figure*}[t]
    \centering
    \begin{minipage}{0.55\textwidth}
        \centering
        \subfloat[F1-Early phase\label{fig:f1_early}]{%
            \includegraphics[width=0.3\linewidth,height=2.5cm]{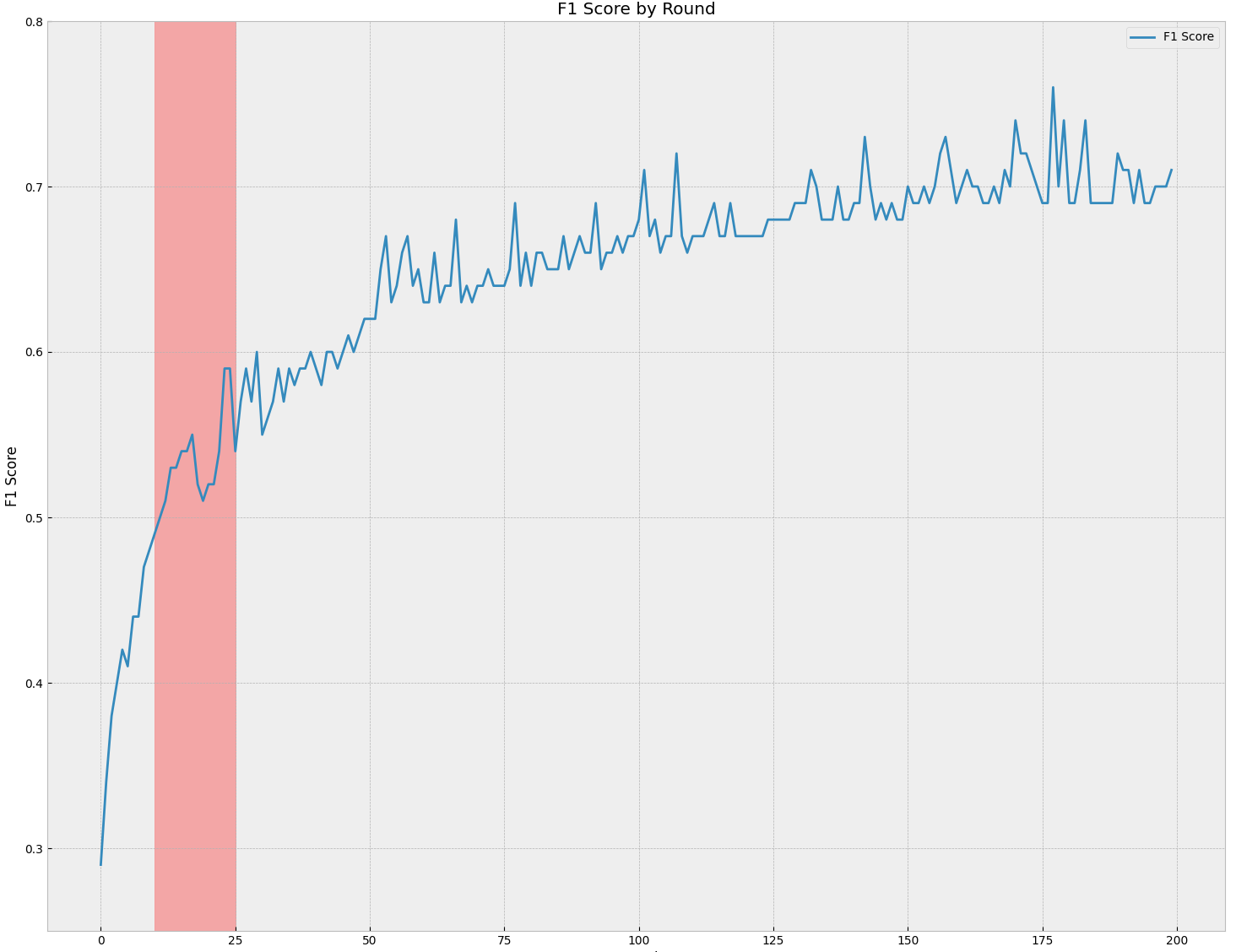}}
        \hspace{0.2cm}
        \subfloat[F1-Mid phase\label{fig:f1_mid}]{%
            \includegraphics[width=0.3\linewidth,height=2.5cm]{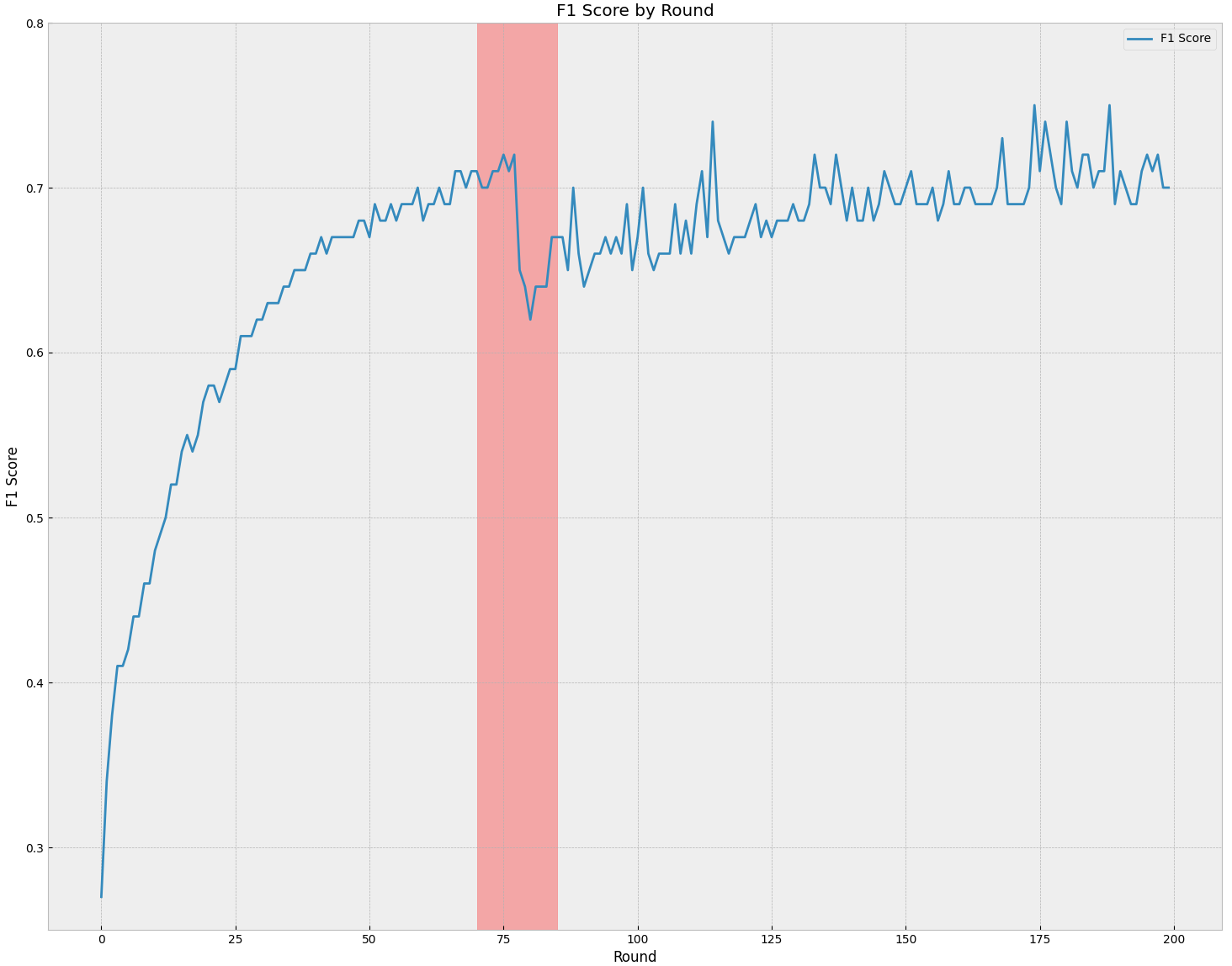}}
        \hspace{0.2cm}
        \subfloat[F1-Late phase\label{fig:f1_late}]{%
            \includegraphics[width=0.3\linewidth,height=2.5cm]{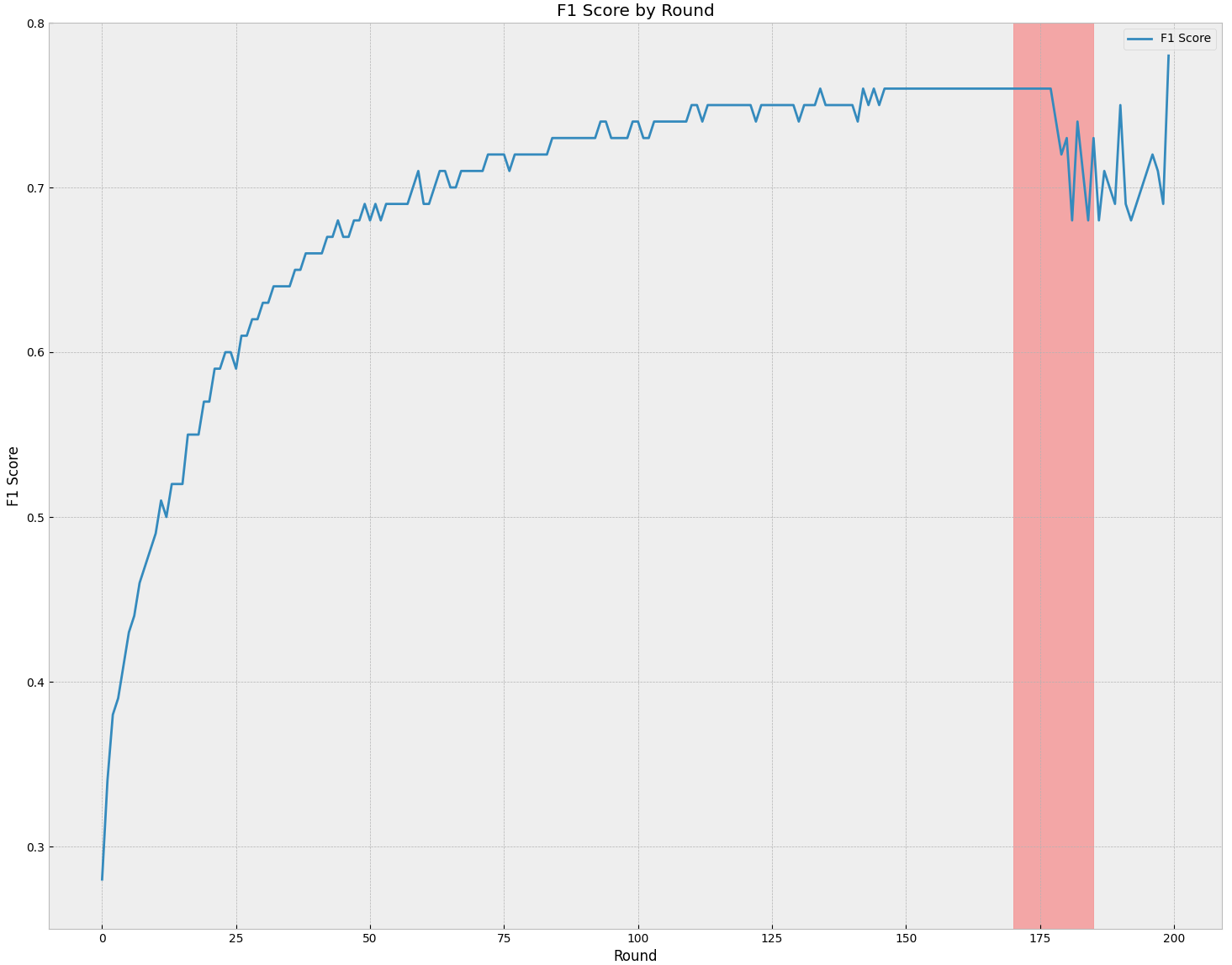}}
        \caption{\small F1 score graph through different phases.}
        \label{fig:f1_score_graph}
    \end{minipage}
    \hspace{0.5cm}
    \begin{minipage}{0.4\textwidth}
        \centering
        \subfloat[5/50 malicious clients \label{fig:less}]{%
            \includegraphics[width=0.47\linewidth,height=2.5cm]{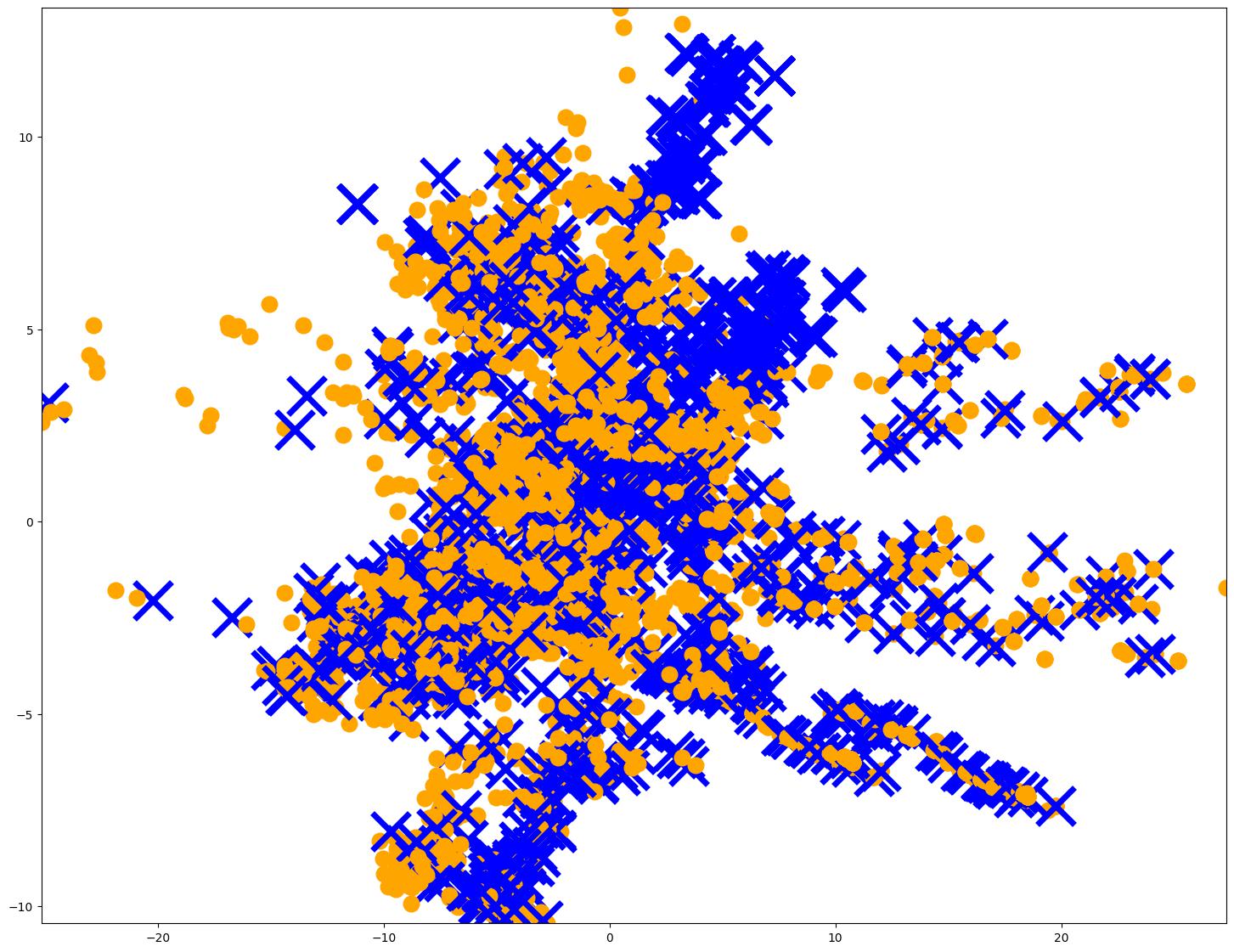}}
        \hspace{0.2cm}
        \subfloat[30/50 malicious clients \label{fig:more}]{%
        \includegraphics[width=0.47\linewidth,height=2.5cm]{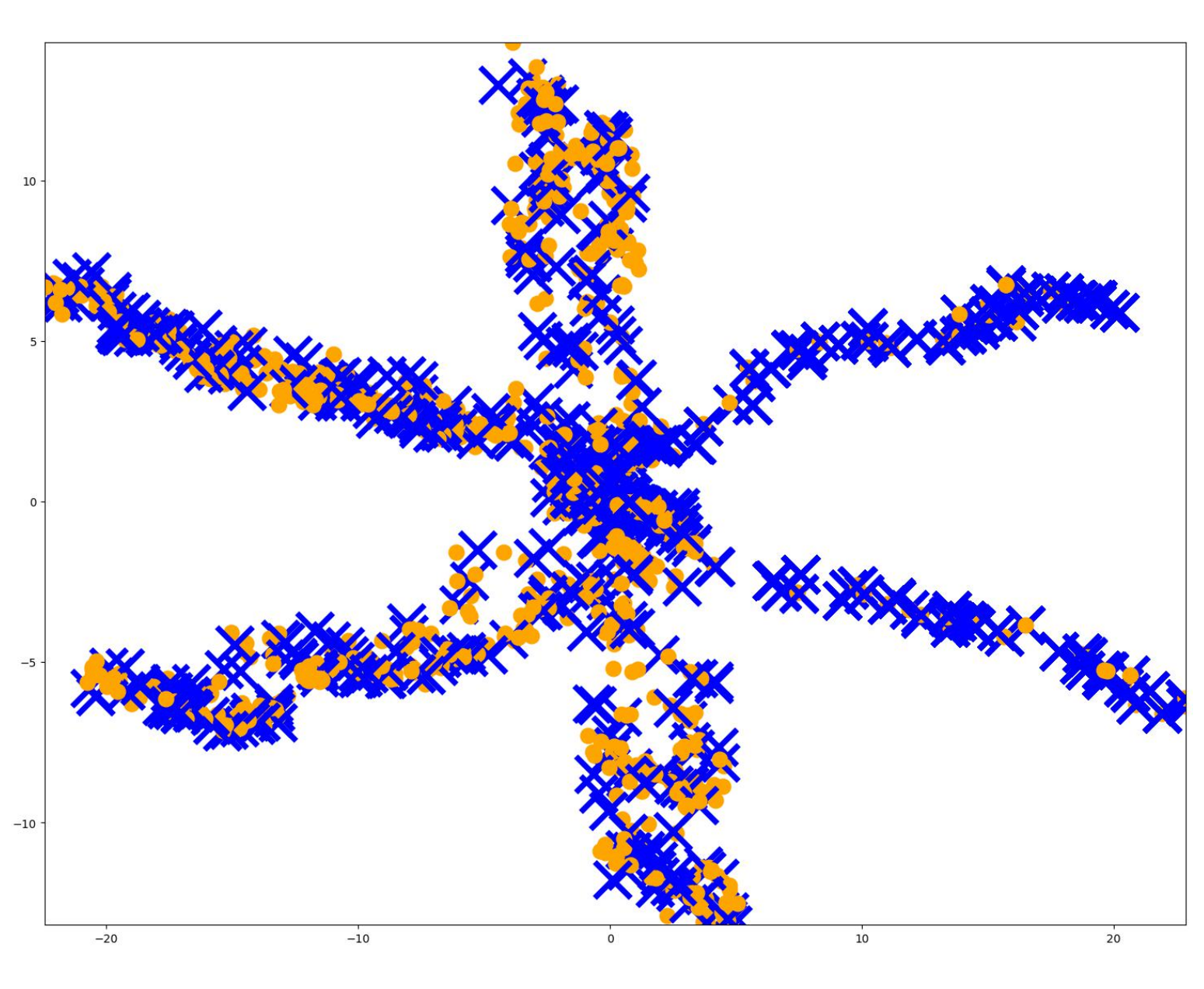}}
        \caption{\small Abnormal PCA effect in one round.}
        \label{fig:abnormal-pca}
    \end{minipage}
    \vspace{-0.4cm}
\end{figure*}

\noindent\textbf{Utility analysis: F1 score.} To investigate the most effective data poisoning strategies, we vary the attack timing and the probability of malicious client participation. We hypothesize that attacks implemented in the late stages of model convergence are potentially more disruptive compared to early interventions. This hypothesis is examined by comparing the impacts of early-round poisoning with later-round attacks. 

Our analysis revealed two key insights regarding the effectiveness of data poisoning attacks. Firstly, the timing of the attack plays a crucial role in maximizing its disruptive impact. Attacks implemented too early in the model's convergence phase tend to be mitigated as the model continues training and recovers from the initial poisoning, as shown in \textbf{Figure~\ref{fig:f1_early}}. Conversely, attacks executed too late may not allow sufficient time for the poisoned model updates to propagate and influence the global model effectively, as shown in Figure~\ref{fig:f1_late}. Our experiments demonstrated that attacks initiated around epoch 70, as shown in Figure~\ref{fig:f1_mid}, approximately one-third into the training process, yielded the most substantial degradation in model performance. Secondly, the intermittent nature of the attacks manifested as fluctuations in the F1 scores for both the target class and the victim classes. This observation suggests that the model's ability to learn and generalize was periodically disrupted by the introduction of poisoned data, leading to oscillations in its predictive performance.

\noindent\textbf{Attack behavior analysis: PCA Signature.}
To visualize the attack behavior on the local model update, we employ PCA with orange dots as benign clients and blue crosses as malicious ones. We have made several interesting observations on the proposed demo. 
\begin{enumerate}
\item \textbf{Existence of separable local model update}: In the presence of data poisoning attacks at local clients, local model updates from malicious and benign clients may demonstrate separable phenomena in the PCA-transformed feature space. This separability could allow for the identification of outlier clusters corresponding to poisoned data.
\item \textbf{Ineffective Poisoning Attempts}: When malicious clients are few, and their poisoning attempts are made early in the training process, the aggregate effect on the global model is minimal. The benign training from subsequent rounds 
weakens the poison effect. 
As shown in \textbf{Figure~\ref{fig:less}} with 5 malicious clients out of 50 total clients, PCA fails to mark any significant deviation from the norm, as the visual clustering of data points remains largely unaffected. When the two cluster merge together, the attack effect is low.
\item \textbf{Overwhelming Poisoning}: When the poisoning is aggressive—signified by a large number of malicious clients—the model begins to learn the biases as features. The two clusters combine, and the impact of the attack diminishes. Figure~\ref{fig:more} shows the result with 30 malicious clients out of 50 total clients. The attackers effectively shift the 'norm' in the feature space, a phenomenon we observe as the model's compromised integrity. These results echo the complication scenarios of FL~\cite{wei2023demystifying} when relying on the separable phenomenon of benign and malicious model updates for outlier detection. 
\item \textbf{Spatial-Temporal analysis}: Besides the spatial analysis of the separable phenomenon on model updates, we further integrate the temporal dimension into the PCA analysis by tracking gradient updates across training epochs. By mapping data points based on both client group similarity and generation round and sequence, we can visualize not just static snapshots but also dynamic changes. This method enables examing dynamics in the model's learning trajectory, revealing the progressive impact of poisoned data. The observation that malicious clients' model updates demonstrate a denser distribution than the benign ones makes it possible to detect the existence of attacks more effectively~\cite{chow2023stdlens}. 

\end{enumerate}

\section{Advisory System}


Through thorough visualization and analysis, we determine the conditions under which model performance is most susceptible to poisoning strategies, employing sensitivity analysis to highlight FL system vulnerabilities. 


\noindent\textbf{Enhanced Client Verification Strategies}: By tracking the confusion matrix on the class-wise performance of the models, client verification processes can be introduced to maintain data integrity. Clients presenting data that significantly lowers the F1 score will be flagged for further scrutiny or excluded from the model update pool.

\noindent\textbf{Anomaly Detection Mechanisms}: Outlier detection via a separable model update from PCA/SVD/clustering can fail to identify malicious clients due to the complication scenarios of FL. Monitoring temporal information with advanced anomaly detection is required to smoothly quarantine malicious participants and undo their negative impact in real-time.

\noindent\textbf{Robust Model Development Practices}: We advocate for models that continuously learn and adapt, using insights from F1 scores and the behavior of local model updates. Incorporating techniques like cross-validation, robust aggregation, and data/model sanitization could potentially minimize the impact of poisoned data.

\noindent\textbf{Robustness Performance Co-Design}:
Apart from the development of robust approaches, it is equally important to maintain model performance~\cite{wei2024trustworthy}. Different aspects of robustness consideration as well as performance guarantee should be embedded right from the design phase.

Leveraging both the utility analysis and attack behavior analysis, FL systems can establish a more informed defense posture against data poisoning. By
intertwining these analytical tools with proactive and reactive
security measures, the FL framework can not only detect but
also adapt to the evolving landscape of data poisoning attacks. 

\section{Conclusion}

This demo paper presents a visualization and analytical platform that simulates targeted data poisoning attacks via
label flipping and analyzes the impact on model performance. The demo system contains five components: Simulation and Data Generation, Data Collection and Upload, User-friendly Interface, Analysis and Insight, and Advisory System. Our demo system provides valuable insights on label manipulation, attack timing, and malicious attacker availability through F1 analysis and signature analysis, offering strategic recommendations toward the robustness of FL systems.

\bibliographystyle{IEEEtran}
\bibliography{demo.bbl}

\end{document}